# A Route to school Informational Intervention for Air Pollution Exposure Reduction

Shiraz Ahmed, Muhammad Adnan*, Davy Janssens, Geert Wets
Transportation Research Institute (IMOB), Diepenbeek, 3590, Belgium

**Abstract**

Walking and cycling are promoted to encourage sustainable travel behavior among children and adults. School children during their travel episode to-and-from school are disproportionately exposed to air pollution due to multiple reasons such as proximity to high traffic roads and peak volumes. Regular use of less polluted routes to and from school can bring significant health benefits for school children. This paper presents a route to school informational intervention that was developed incorporating approaches and methods suggested in the literature for effective behavioral interventions. The intervention was implemented using escorting parents/guardians (*N*=104) of school children of Antwerp, Belgium to adopt school routes with least exposure to pollutants. Collected data and its analysis revealed that 60% participants (N= 62) could benefit themselves by adopting the suggested cleanest routes to school, of whom a significant proportion of participants (i.e. 34%, N= 35) have a difference of average $NO_2$ concentration between the alternative and current route of around $10\mu g/m^3$. This information about alternatives routes with their potential benefits was presented to each participant via defined study protocols. Based on the feedback of participants that could potentially adopt suggested alternatives, 77% (N=48) have switched their routes. These results indicated that intervention was effective, and it can bring higher benefits when implemented on a wider scale.

**Keywords**: Children, routes to and from school, traffic, exposure, air quality, informational intervention,

## 1. Introduction

Good air quality is essential for the healthy living of human beings. Despite that, more than 80% of the people living in urban areas of the European region are exposed to higher levels of air pollutant concentrations than the World Health Organization (WHO) limits [1]. In the last decade, an insignificant reduction in average pollutant concentrations has been observed in the European Union (EU). In Europe, pollution and reduces life expectancy by almost 9 months [2]. It is evident from the latest findings of WHO that children are most susceptible to the severe impacts of air pollution [3]. Even ambient air pollution has an effect on child health before birth if the mother is exposed to high levels of pollutant concentrations [4]. According to WHO statistics, 93% of all the children are exposed to pollutant concentration levels above the limit values. More than one in every four deaths of children under 5 years is directly or indirectly associated with environmental hazards [5]. In 2016, ambient air pollution causes respiratory diseases that resulted in 543,000 and 52,000 deaths in children under 5 and 5 to 15 years respectively [3].

In Europe, air pollution has mostly resulted from personal motorized vehicles [6]. It has found that exposure to traffic-related pollutants primarily $NO_2$ impact on children lung development and cause air passage disorders such as inflammation, chronic bronchitis, sensitization and allergies [7]. Since children are growing and their immune system are still developing, so they are particularly more affected by the pollutant exposure [8]. Pedrerol et al [9] has mentioned that children are highly exposed to traffic-related air pollution while their home to school commuting due to high traffic pollution peaks. It has highlighted in the studies that even though time spent during school commuting might be less but children are exposed to a high proportion of their daily pollutant dose during that episode of travel [10-12]. BREATHE project has found that children on an average spent 6% of the daytime in commuting that resulted in nearly 20% of the daily integrated dose of Black Carbon (BC) [12]. On the other hand, active school travel (AST) (i.e. walking and cycling to school) has been encouraged and promoted worldwide using different intervention strategies [13-21]. However, high respiratory rate and enhanced level of physical activity during walking and cycling may result in a high inhaled dose of air pollutant. In a study for Manchester (UK) [22], primary school children walking routes were simulated considering school and household locations. The routes were simulated using either the shortest duration or lowest cumulative $NO_2$ or $PM_{10}$ concentration that a child could be exposed. The study found that for at least 50% routes, walking along alternative routes which are 2 minutes longer than faster routes resulted in 5 µg/m$^3$ reduction in $NO_2$ exposure. Routes to school that children follow must have pollution concentration below the specified limits. Escorting parents, school administration, public health monitoring agencies and relevant health advisors are, therefore responsible for acquainting themselves with such information and then communicate accordingly. So that highly polluted routes are avoided. The study for Manchester (UK) was exploratory, therefore it is required to take one step further. This paper aimed at engaging escorting parents/guardians and made them aware of such information to influence the selection of a better alternative route to school.

There are a variety of ways mentioned in the literature through which intervention can be carried out by providing customized information at an individual level. For example; providing feedback to individuals about their current behavior in terms of their consequences, giving justification regarding possible changes in behavior using appropriate quantification of benefits it could bring, taking their commitments and setting goals to bring a desired change in a particular behavior and incentivize individuals to bring desired changes in the behavior [23]. These methods have tested specifically about recycling and energy conservation all over the world, but there are hand full of studies conducted to influence and change travel behavior [24]. In Columbia (USA), bicycle proficiency program in a classroom environment was launched, and it was found that 35% of the car trips were replaced with bicycle trips in a week [25]. Another study based on individualized marketing strategy was conducted in Perth (Australia) in which 35000 people participated. This strategy resulted in positive travel behavior change, i.e. increase of 35 %, 61 % and 9 % were noted in walking, cycling and traveling as a car passenger respectively [26]. Cairns et al. mentioned that improvement in public transport services coupled with efficient informational campaign resulted in increased bus journeys by around 42% [27]. Similar trends have been seen for an informational campaign (named as indiMARK®) to promote public transport in Germany, Australia and UK. Customized information about public transportation, cycling, and walking was provided to individuals that include guides to walking and cycling in the locality, schedules, and maps for public transport routes, consistent with individual requirements [28].

Schultz [29] mentioned that for higher success and impacts interventions should be designed as such that they are focused on aspects of individual travel behavior which are comparatively easier to change and at the same time bring high benefits. Easy to change travel behavior can be for example; the adoption of a green alternative route if the travel time difference is not significantly larger from the current route, replacement of short car trips (i.e. travelling distance within 3 km) to cycling when there is no time pressure and other constraints. Adopting such changes results in benefits such as improve health and reduce traveling cost [30]. Efforts were made to quantify the possible benefits of replacing car-based short trips by active modes and public transport. Beckx et al [32] for Dutch car drivers assessed the number of short car trips that can be replaceable by active modes. They found that 10% of car trips were replaceable that resulted in savings of 3% fuel consumption. In another study, Fellerman found that a particular car trip, if replaced with public transport, can reduce 140 g $CO_2$, 0.9 g carbon monoxide, 0.17 g VOC, 0.3 g $NO_2$ and 0.008 g PM per km [33]. These traffic-related pollutants harm children health. Kim found that respiratory symptoms are more elevated in children living in close proximity to traffic. Strong associations were found between traffic-related pollutants and asthma and bronchitis symptoms in a highly urbanized region of the United States with good regional air quality, where local air pollution is dominated by vehicular sources [34]. As children are especially more sensitive to the adverse effects of air traffic-related air pollutants, therefore home to school routes with lower pollution levels could provide higher health benefits for them.

Informational interventions designed and implemented so far are based on strategies to educate, motivate and encourage individuals by providing general information to promote pro-environmental travel behavior. There is no such structured and organized effort found in which information is provided considering the context of an individual with a detailed illustration of the consequences of such behavior and benefits of suggested changes in behavior. For example, strategies are not based on examining their details of travel behavior, so that focus should be on improving a specific part of their travel behavior. Such as providing individual specific guidance about safer and lesser pollutant routes to school with the benefits of suggested changes. It is important that in the intervention programs, this aspect is highlighted more clearly so that children/escorting parents are more inclined to follow such routes which are potentially safer in relation to pollutant exposure.

The objective of this study was to find out the replaceable potential by examining children's current route to school pollutant exposure and availability of lesser polluted routes. Furthermore, exploitation of that replaceable potential by setting up an informational intervention and assess that up to what extent such intervention is effective. The intervention has been set up primarily by providing customized information about the consequence of current and suggested school travel choices concerning the following aspects.

- Exposure to air pollution based on current school travel choices
- Details such as distance, travel time etc. about available walking/cycling school routes
- Exposure to air pollutant based on suggested walking/cycling school routes
- Benefits of proposed route changes

The details regarding the design, implementation and effectiveness of this informational intervention are well explained in the following sections. This paper is structured as follows: Section 2 explains the data collection techniques adopted for gathering required data sets along with data integration and processing to get the desired information. Section 3 presents details of the implementation of the designed intervention. Section 4 discusses the obtained results of the study. Section 5 provided a discussion on the results along with practical implications followed by the conclusion section.

## 2. Materials and methods

This section provides details about the required data sets along with their acquisitions methods. Based on these acquired data sets, a distinct mechanism was developed to generate easy to adopt school travel choices keeping in view the air quality. The generated output was used to set up information based behavioral intervention to influence changes in the route to school. Further insights about each of the steps involved are mentioned below.

*2.1. Data Required and acquisition methods*

A range of dataset was acquired to detect current and alternative home to school routes along with their pollutant exposure level. These data sets comprised of not only detailed information taken from the escorting parents or children regarding home to school routes but also about the pollutant concentrations in that microenvironment. Furthermore, additional information such as availability of walking and cycling routes along with other associated features were also acquired and stored as third party dataset. The paragraphs below provide details about how these datasets are acquired.

Children school route information was recorded using Route2School online platform developed in house by Transportation Research Institute (IMOB) at Hasselt University. This online platform can be accessible both using mobile (https://play.google.com/store/apps/details?id=com.dhcollator.routetoschool) and web application (www.route2school/login). This user-friendly online platform shown in Figure 1 was provided to the escorting parents/guardians, where they can easily record their children's home to school information. Parents/guardian fed in some initial information such as home and school address along with the usual travel mode. Based on this, an appropriate route was displayed on the GUI of OpenStreetMap. Parents/guardians could adjust the route by dragging it on the currently used route if required, so that recorded route have the right combination of road links.

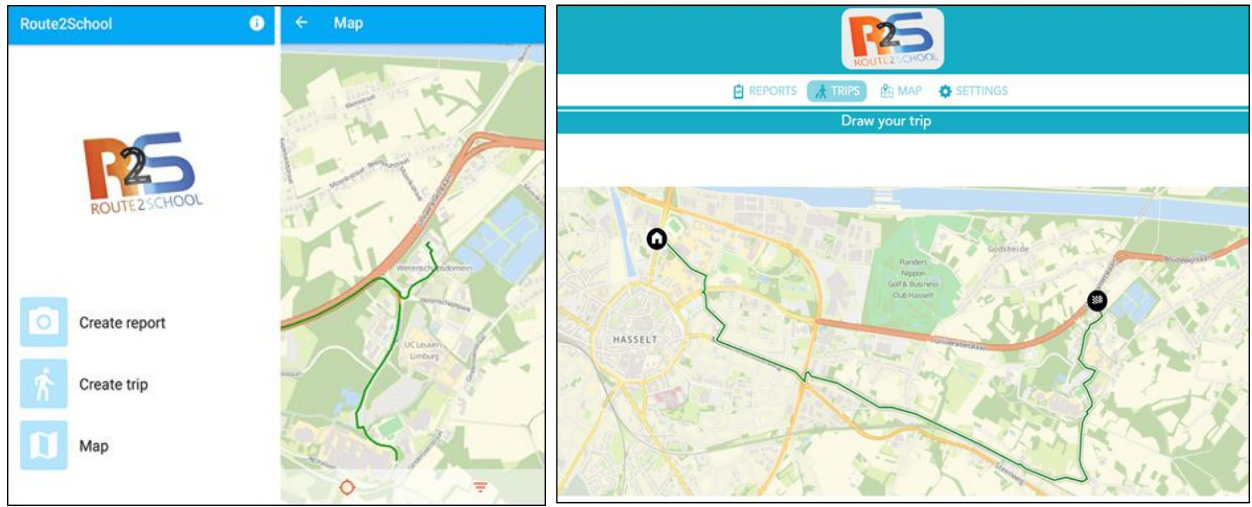

Figure 1: Route2School mobile application (left) and web application (Right) user interface

Children are highly exposed to on-street traffic pollutants while commuting along busy roads [35]. In the Flanders region of Belgium, about 50 % of the total $NO_2$ emission is caused by road traffic. As such, $NO_2$ is an important marker for quantifying traffic-related air pollution [36]. According to WHO, children are more vulnerable to the impacts of $NO_2$ compared to other pollutants [3]. $NO_2$ concentration data at fine spatial resolution is required to get insights into child exposure while school commuting. So in this regard, hourly annual $NO_2$ concentration at a spatial resolution of 10 m by 10 m grid map is obtained for Flanders from Flemish Institute of Technological Research (VITO). This map is depicted in Figure 2. It is based on the measurements recorded from various stations in Flanders and the surrounding regions, using interpolation and high-resolution modeling. Different concentration levels of $NO_2$ are shown using green, light green, yellow, red and purple colors from the lowest to highest concentration levels respectively. ATMO-Street (RIO-IFDM-OSPM) model is used in the estimation of hourly $NO_2$ concentration values at street level. This holistic model is based on the integration of the following models [37].

- RIO is a land use regression model for the interpolation of hourly pollutant concentrations as measured by the monitoring stations

- The IFDM (Immission Frequency Distribution Model) is a bi-Gaussian plume model used for calculation of the air quality based on meteorological data and the emission of air pollutants
- OSPM model contains detail information about the street configuration and incorporates the impact of street canyon effect while simulating street-level concentrations.

The advantage of the RIO-IFDM-OSPM combination is that the air quality can be estimated with a high spatial resolution down to street level. The validation analysis was performed at various stages in the model chain, explaining the strength of different components for the development of concentration maps. The model represents the spatial variability at urban to street level scale with an $R^2$ value of 0.86 [38]. A recent study conducted in Milan, Italy predicted cleanest (having less black carbon exposure) home to school routes using the LUR model [39]. For 2 months, children BC exposure on home to school routes were also measured using personal monitoring kits. The exposure estimated using kits and LUR model showed good agreement having Pearson's r = 0.74, Lin's Concordance Correlation Coefficient = 0.6. The results indicated that the prediction of cleanest school routes can be done using LUR estimates.

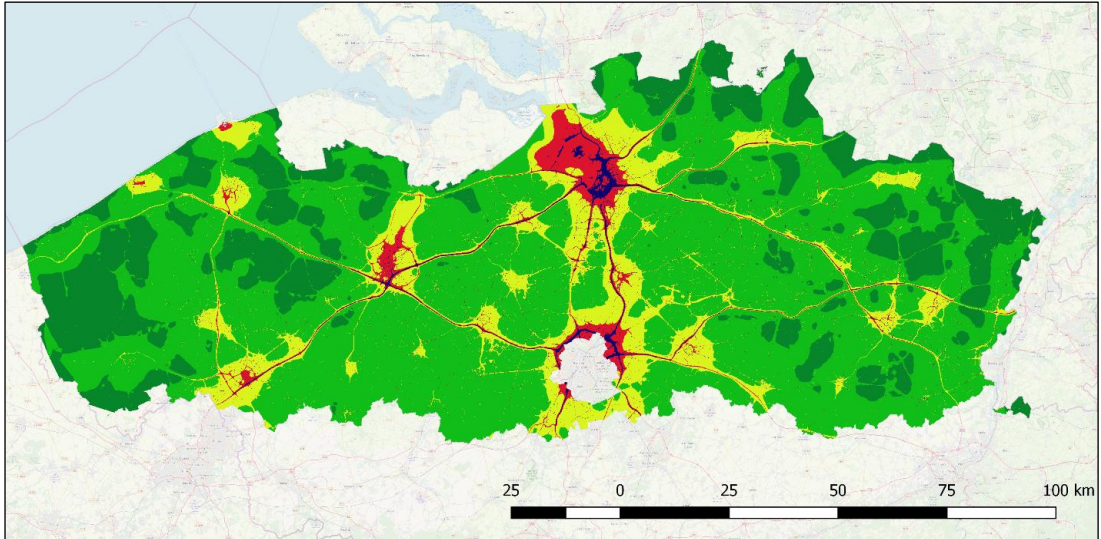

Figure 2: Map showing 9$^{th}$ hour (08:00 am) average annual NO$_2$ concentrations of Flanders for school days

Data about infrastructure related to cycling and walking is also required in computing alternative routes based on active commuting. This information was obtained from available route planner platforms such as Graph Hopper Directions API in which routes are optimized with the help of OpenStreetMaps. Details associated with each route such as gradient, number of intersections, availability of segregated path etc. could also be found using this API [40].

*2.2. Rules-alternatives active school routes*

Rules to detect alternative walking/cycling routes to school are based on the threshold limits and constraints involved in children school commuting. These threshold limits and constraints were identified from the contextual information and past studies done in this regard. In one such study threshold distance of 3 km from home to school is suggested to promote active school travel in Flanders [41]. The rules to identify the most feasible route alternatives are as follows:

- Alternative cycling routes must be ≤ 3 km if the current route is ≤ 2.5 km
- Alternative walking routes must be ≤ 1.25 km if the current route is ≤ 1 km
- The difference of alternative to current route must be ≤ 0.25 km if the current route is ≤ 3 km.

- The difference of alternative to current route must be ≤ 0.5 km if the current route is ≥ 3 km.
- In case of walking, footpaths are available all along the route
- In case of cycling, segregated bike lanes are available at least halfway along the route
- Crossings in the bike route/footpath are limited to 3
- Route average gradient is not more than 10%

*2.3. Data Processing*

The mechanism adopted to detect less polluted walking/cycling routes to school involves various processes and computations. It starts with retrieving recorded home to school routes from R2S platform followed by identifying feasible walking/cycling alternative routes and ends by evaluating pollutant exposure involved per trip. Further, the information generated was organized and structured in a way that it is easily incorporated in the intervention tool. The details about how various data sets integrate during different processes and how they flow from one process to another are explained below and depicted in Figure 3.

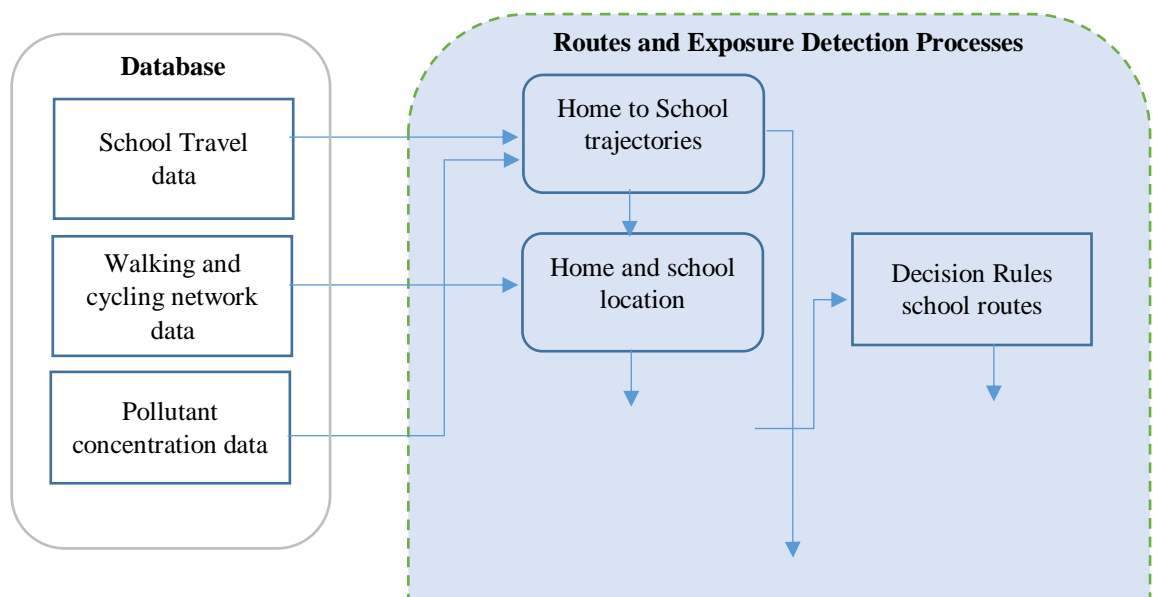

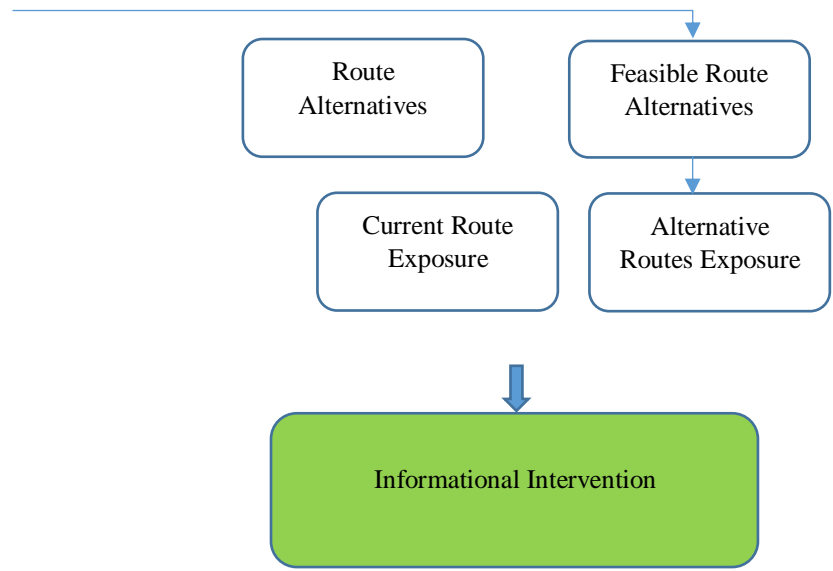

Figure 3: Flow Diagram for the assessment of less air polluted alternative school routes

Initially, all recorded routes were retrieved from R2S platform and stored in the database as a single file in .dbf format. Each route has certain attributes such as project id, route id, home and school location and mode used. The home and school location are used as one of the inputs to detect alternative walking/cycling routes. For this purpose, an existing algorithm could be used such as k-shortest path approach, link elimination and simulation methods [42]. Depending upon the simple input data requirement, the k-shortest path approach was selected. In the 1st phase, the process was initiated by executing k-shorted path algorithm that requires cycling /walking network data (obtained from ordinary street maps for Flanders region of Belgium) and origin-destination details (x,y pairs) as input. The resulting output was stored as "Alternatives Routes". The detected route alternatives were screened further based on the rules mentioned in section 2.2 and stored as "Feasible Alternative Routes". So the developed algorithm identifies such school travel route alternatives that are relatively easy to be adopted to fulfil the intervention effectiveness requirement mentioned by Schultz [28].

In the 2nd phase, pollutant exposure was computed for current and alternatives routes. Pollutant concentration maps of hourly annual $NO_2$ concentrations at 10 $m^2$ grid resolution was stored as

.geotiff format in the database. Exposure detection program processed route trajectories and generated (x,y) pairs at an interval of 10 meters to avoid resolution mismatch. This was done using line to point convert module in QGIS. Each unique (x,y) pair of the specific route was assigned the $NO_2$ value from the concentration map using the nearest neighbor approach. The resulted output was stored as "Route points with pollutant concentration". In the end, each point of the route has an attribute of "pollutant concentration" value as shown in Figure 4.

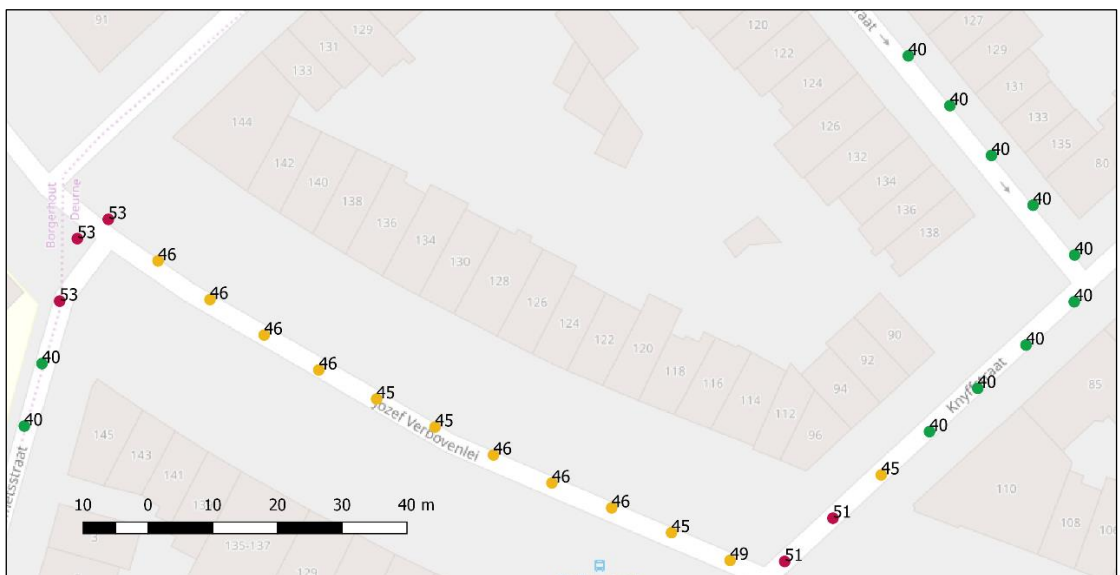

Figure 4: Points of the route with $NO_2$ concentrations values

These data integration and processing phases finished with alternative active school routes with pollutant concentration level per route. The per route average concentration was estimated using a weighted average formula based on the total number of data points for each route. For each individual, alternative routes were further compared and ranked from lower to higher average concentration values for the development of intervention information package.

*2.4. Intervention tool: Customized Information Package*

Mobility based informational interventions have a pull effect and ability to effect socio-cognitive aspects of individual travel behavior by applying strategies based on information dissemination and persuasion [43]. Therefore the efficacy of such an intervention primarily depends on the ways how the information is designed. In past studies, various methods were adopted and examined for their effectiveness [24]. Based on the review by Adnan et al [44] it was found that an informational interventions based on the following approaches were more appropriate to influence change in travel behavior. A brief explanation of these approaches are as follows:

- *Feedback*: monitoring individuals, and then providing feedback regarding their travel behavior such as how much is their exposure to air pollutant for their activity-travel routine.
- *Justification*: Reasons can be provided about the recommendation of a particular change in behavior such as if short trips are made via active travel modes then what are their benefits.
- *Instructions*: Procedural information is provided to carry out specific behavior; such as avoid car use in short trips within 3-5 km.

Keeping in view the above approaches, an online customized information package was designed that act as an intervention tool to influence the change in school travel behavior. The overall effectiveness of the customized information package also significantly depends on the structuring i.e. organization and presentation of the information. So each section of the Customized information package is organized as follows:

- *Contextual information* with infographics to increase the awareness related to air quality and pollutant exposure impacts, which are easy to understand and digest.
- *Customized Feedback* regarding a quantitative measure of current and alternative school travel choices with the description of their impacts in terms of pollutant exposure.
- *Description of the personal benefits* achieved in terms of reduction in pollutant exposure level by adopting an alternative route and long term health impacts.
- G*eneral suggestions and tips* on how to avoid pollutant exposure in an urban environment.

Further to avoid any confusion, the routes of the detected alternatives were marked on the GUI of google street maps and embedded in the Customized Information Package as shown in Figure 5 below.

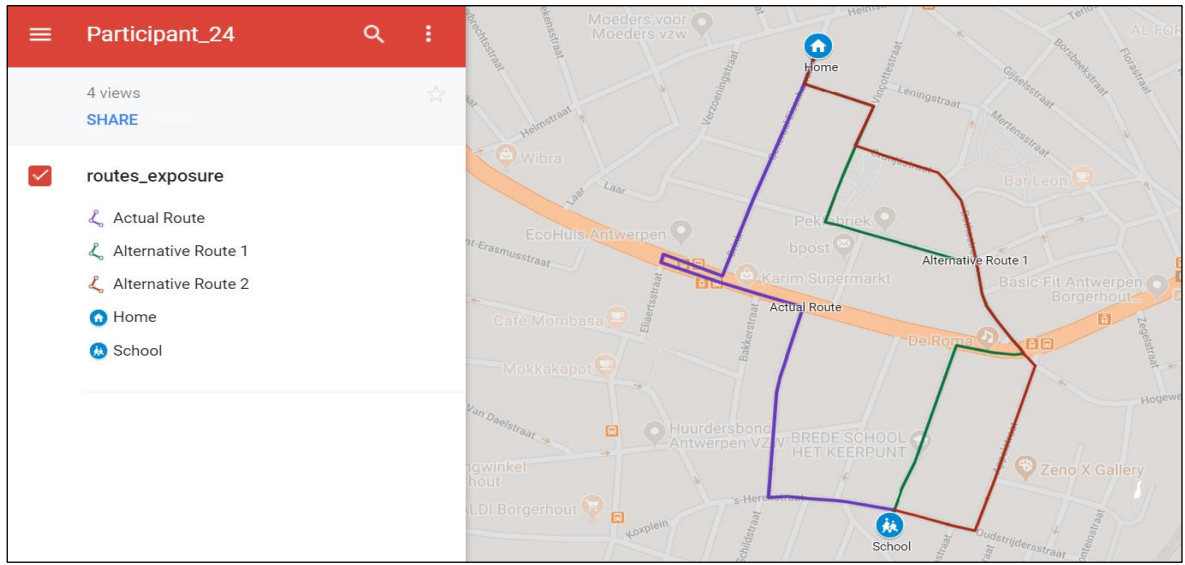

Figure 5: View of embedded routes in online Customized Information Package

## 3. Intervention Study Details

### 3.1 Study Area context

The study area consists of Berchem and Borgerhout region of Antwerp city as shown in Figure 6. The city has a central location in a highly urbanized region in Europe. Since it is an important port city and transportation hub, the commuter and the international transit traffic are also part of the traffic flows in and around the city [45]. The Antwerp ring road carries most of the traffic in this region. On average 300,000 vehicles use this road link daily in which 27 % are freight vehicles. The Antwerp ring road is passing through areas such as Borgerhout with population densities above 11,000 inhabitants/km². Around 352,000 people reside within 1.5 km radius from the ring road [46]. 12 schools and 75 nurseries are located in the nearby area, where daily average particulate concentrations were observed above EU threshold limits. As a result, it is a recognized polluted hotspot, particularly $NO_2$, which has adverse respiratory effects, especially in children [47]. This background information is a major motivation factor to carry out this study.

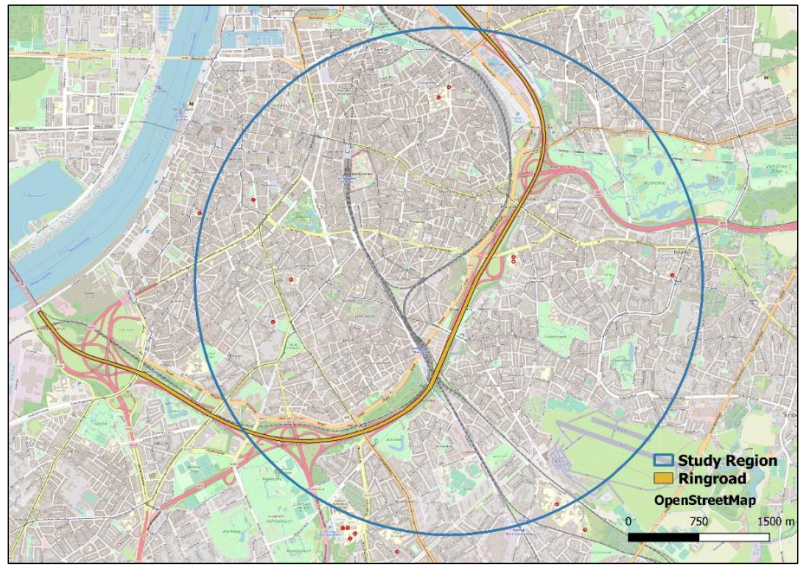

Figure 6: Figure showing localities along Ring road within study area

*3.2 Recruitment Process*

Before the launch of a campaign for the participant recruitment, ethical approval was taken from the Social-Societal Ethics Committee (SSEC) of Hasselt University. Information about the study was floated using social and local media. It was mentioned that the study target group should belong to the surrounding areas of the Antwerp ring road, who are interested to know less polluted school routes for their kids. 113 interested citizens, i.e. parents of elementary school kids, contacted us via email in the month of February 2019. They were asked to join the study by signing an electronic informed consent form. There were some dropouts as well, and 104 participants managed to complete the study successfully. It was observed that 85% of the parents escorted their kids to and from schools. The overall methodology of the study was designed in such a way that it can be managed remotely, without the requirement of physical contact. Therefore, citizens who are participating in the study needed to have sufficient knowledge of e-mail based communication and some know-how of web-based tools.

*3.3 Implementation Procedure*

Implementation of the study was done in four stages. At the start of the study, participants were asked to fill in the online introductory questionnaire by sending them an online link via email. This task aimed to know about the individual (escorting person/guardian) socio-demographic details and their awareness about air pollution. In the $2^{nd}$ stage, participant current school travel behavior was recorded. The detailed information about how to access and work with Route2school online platform was sent via email. Escorting person/guardian marked home to school route with required attributes and sent the information to the database. More than one route was recorded if different routes were followed at the start and pack up time or different days of the week. Recorded information about current school behavior is processed and analyzed, as mentioned in section 2.2. Participants were given feedback in the form of online Customized Information Package. This package provided insights about actual and suggested school travel choices along with their consequences. The layout of the electronic version of this package is shown in figure 7. After the week, participants were asked to fill in the feedback questionnaire to know about the possible effect of information provided and generic comments about the study. The whole procedure is well depicted in Figure 8.

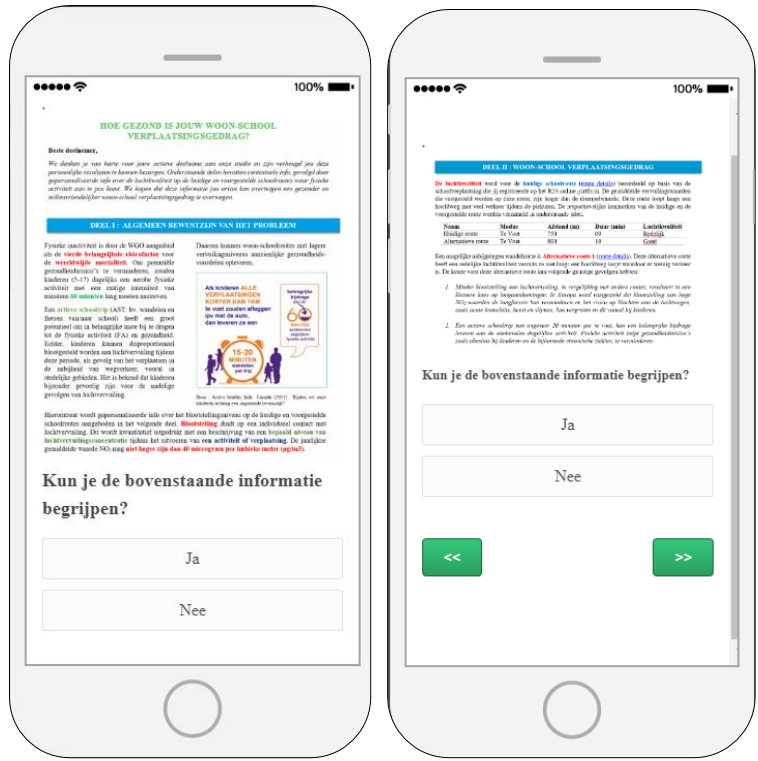

Figure 7: Layout of the Information Package (Dutch version)

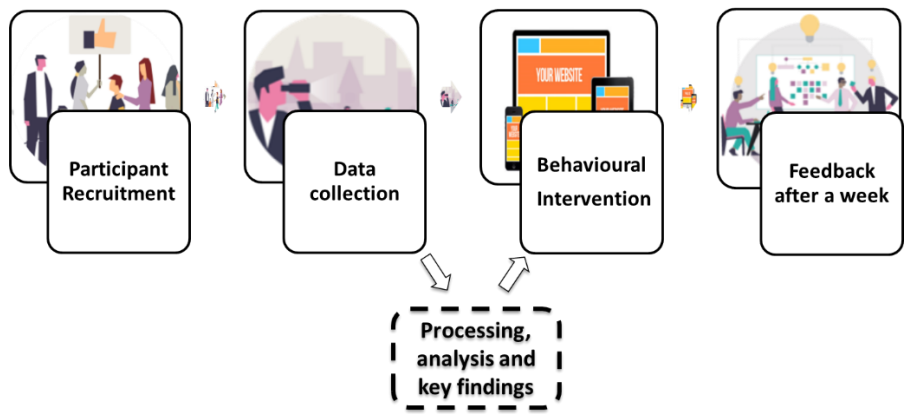

Figure 8: Procedure involved in the implementation of the study

## 4. Results

This section highlights the main findings of the study implemented in the Antwerp region. Analysis of different data sets collected during the study is presented in detail. Along with the descriptive and qualitative analysis of the contextual data, current and suggested school travel choice were also estimated that provide key results on which information intervention study was based. Furthermore, effectiveness of the air quality based information intervention is also presented.

*4.1 Descriptive analysis*

This section highlights key socio-demographic characteristics and views of the study participants about school commuting and air pollution. Furthermore, some situational factors such as walking and cycling infrastructure details were also gathered. Table 1 below illustrates the distribution of socio-demographic characteristics of the participants involved in this study.

Table 1 provides interesting details, it can be noted that the majority of the participants in terms of the 2 age groups, i.e. between 30–50 years old have participated in the study. Female participation is at the higher side compared to males. Majority of participants have education status equivalent to bachelor's degree or above (i.e.79 %), and income distribution is similar to a typical negatively skewed with a median value around 35,000 € per annum. According to Eurostat [58], monthly minimum salary of a Belgian resident for year 2019 is 1593.81 euros which is around 19,000 euros per annum. That is the reason that there is a least probability that participants fall in the income category less than 10,000 euro per annum. Keeping in view the sociodemographic of the Belgians, the sample included participants from all classes. It can also be observed from Table 1 that most of the families have a size not more than 4 members and the age of the kids who are escorted to school are nearly uniformity distributed among three age categories. Further, around 97 % of citizens have ownership of a car in which 33 % possess 2 cars, but at the same time, it is also found that all study participants have bicycle ownership as well.

Indicators of environmental awareness were also gathered. Based on the distribution of participant responses, it can be concluded that majority of the respondents have an understanding of the environmental issues and they agreed that humans should take responsibility to improve the environmental condition. It is noted that around 61% of the citizens have the opinion that environmental pollution is a problem in their locality. Respondents are also well aware of the significance of physical activity for children by showing 85 % agreement to such a statement. Regarding the active mobility infrastructure, participants showed mixed views about the facilities available to adopt active travel options such as bicycle and pedestrian facilities. 79 % of the citizens have the view that car use should be more restricted by taking some hard policy measures. It depicts the situation that there is more room to improve the facilities that encourage healthy and active travel behavior among individuals in the study area.

Table 1: Socio-Demographic characteristics distribution of study Participants

| Characteristics | Statistics with description |
|---|---|
| Gender (escorting parents/guardians) | • Male    34 %<br>• Female    66 % |
| Age escorting parent / guardian | No of escorting parents expressed in %age with respect to age group is summarized as follows:<br>• B/W 30 - 40 years    39.5 %<br>• B/W 41 - 50 years    51.5 %<br>• B/W 51 - 60 years    3.0 %<br>• Above 60 years    6.0 % |
| Employment status | • Full time work    54.6 %<br>• Part time work    21.2 %<br>• Self-employed/business    12.1 %<br>• Retired    6.0 %<br>• Others    6.1 % |
| Qualification level | • Bachelor diploma    15.2 %<br>• Master's degree    45.5 %<br>• Professional degree    15.2 %<br>• Doctorate Degree    3.0 %<br>• High school graduate, diploma or the equivalent    15.2 %<br>• Trade/technical/vocational training    6.1 % |
| Household income | Participants are categorized in %age w.r.t total household income per annum as:<br>• 10,000 to 24,999 €  12.0 %<br>• 25,000 to 49,999 €  33.2 %<br>• 50,000 to 74,999 €  27.3 %<br>• 75,000 to 99,999 €  15.2 %<br>• 1,00,000 to 1,49,999 €  9.1 %<br>• 1,49,000 to 1,99,999 €  3.2 % |
| Total family members | Participants are grouped with respect to family size (expressed in numbers) as:<br>• 02 members  12.12 %<br>• 03 members  33.33 %<br>• 04 members  33.33 %<br>• 05 members  18.18 %<br>• 06 members  3.03 % |
| Age of kids | No of kids expressed in %age with respect to age group are as follows:<br>• 3-5 years  30.4 %<br>• 6-8 years  33.3 %<br>• 9-12 years  36.3 % |

*4.2 Travel Routes and their Exposure levels*

Children's school travel behavior was recorded using Route2school online platform. Figure 6 is showing localities of Berchem and Borgerhout along the ring road within the study area. As these localities are in close proximity to the ring road, so residents are highly influenced by the traffic-related pollutants, specifically the weaker group such as children. To reduce this impact easy to adopt alternative walking/cycling route choices with least exposure are identified for influencing the travel behavior of escorting parents/guardians and children.

The route detection algorithm is used to identify possible cycling and walking route alternatives with related attributes such as travel time and distance. Subsequently, the current and alternatives routes were processed further to get exposure per route, as discussed in section 2.2. The exposure associated with current and most suitable alternative (having the least exposure) routes, for a study group, is presented in figure 9 as a box plot. This way of graphical illustration is used to compare the exposure distribution for current and alternative routes for study participants. Based on the comparison of box plots, it can be seen that the overall distribution pattern looks similar as both have almost the same range. However, there are noticeable differences in quartile and median values as alternative routes have lower values. The comparison between the 2 box plots suggests that the alternative routes have the potential to reduce $NO_2$ exposure on children while school commuting.

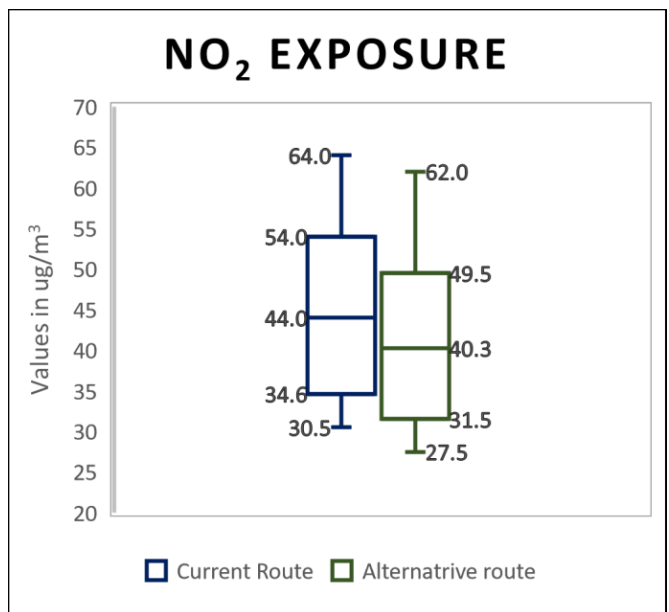

Figure 9: Box plots representing current and alternative route exposure

In order to better communicate escorting parents/guardians about the exposure to different levels of $NO_2$ concentrations, we adopted a categorization based on the limiting values set by WHO and European Environmental Agency (EEA) [3,48]. Three categories are defined, i.e. Low, moderate and high, to characterize the exposure of current and alternative routes. If the average concentration value is less than or equal to 40 ug/m$^3$, the route is categorized as having low exposure, moderate if values are 41 – 50 ug/m$^3$ and high if values are greater than 50 ug/m$^3$. Figure 10 is showing an overview of the exposure variation in the participant current routes. It is found that almost one-third of the routes fall in each category. Around 67% of the participants follow the routes that have exposure values in moderate and high categories, which depicts that every 2 out of 3 children are exposed to concentration values above the threshold limits. The situation demands actions from the authorities. However, in the short term, the exposure could be reduced by encouraging escorting parents/guardian to adopt pro-environmental route choices.

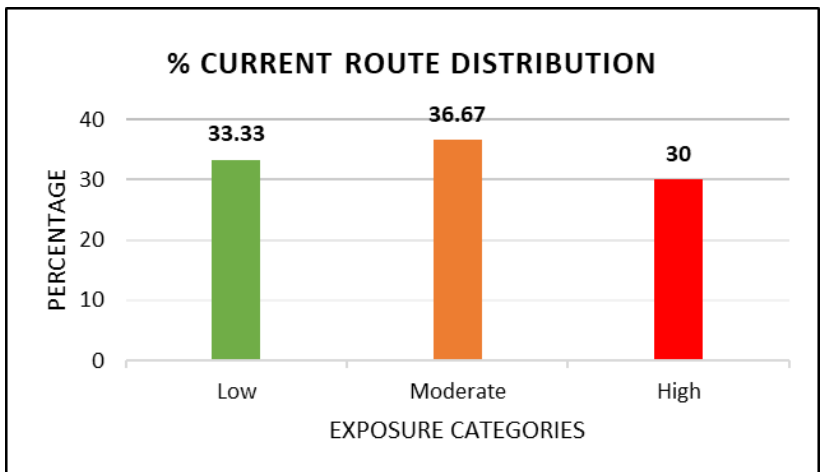

Figure 10: Graph showing actual route distribution based on exposure categories

It is also important to know whether there is any positive difference between exposure of current and alternative route for a child and escorting parent? For alternative routes, exposure categories are assigned similarly and compared to know about the possible benefits in relation to exposure if escorting parents/guardian may choose an alternative route. The results of this comparison are presented in Table 2. The number in each cell in table 2 is representing the %age of the participants whose exposure category can possibly be changed, such as moderate to low, high to low etc. About 17 % of the participant's exposure category changed from moderate (current route) to low (alternative route). Similarly, for 7 and 10 % participant's exposure category can be shifted from high to low and high to moderate category. Therefore, it can be concluded that the significant proportion, i.e. 34 % of the participants has alternative routes with better air quality as compared to their current choice of route. This potential route shift can significantly help to avoid health risks associated with high $NO_2$ concentrations.

Table 2: Participant's exposure category change from current to alternative routes

| % Participant's exposure category shift from actual to alternative routes | | | |
|---|---|---|---|
| Alternative Routes | Current Routes | | |
| | Low | Moderate | High |
| Low | 30 | 17 | 7 |
| Moderate | 3 | 20 | 10 |
| High | 0 | 0 | 13 |
| Grand Total | 33 | 37 | 30 |

Furthermore, the numbers in Table 2 also depict no change in pollutant exposure category for one-third participants (i.e. the numbers 20 and 13 in the table). We also found out the exact differences in exposure values between the current and alternative routes within each category, to see if there are still some benefits exist if participant adopts alternatives routes (though their exposure category remains same after shifting). It is found that there are some significant differences, i.e. reduction in concentration values of alternative routes, as compared to current routes in each category (See Figure 11). The average reduction for Low category is found to be 1.11 µg/m$^3$ with minimum and maximum values of -1 and 5, respectively. A negative value of -1 is showing that some participants don't have any better route as compared to their current route. The mean value of concentration difference between current and alternative routes is 0.75 µg/m$^3$ (-4 to 6) for a participant lie in the moderate category. Similarly, the mean positive difference of 6.25 µg/m$^3$ is recorded for the participants fall in the High category with extreme values of -3 to 10. It can be seen that the average difference for High category is more as compared to low and moderate. The higher the reduction in pollutant concentration values while commuting there is less health risk involved. This link has been statistically established by researchers in various epidemiologic studies [49-51] .

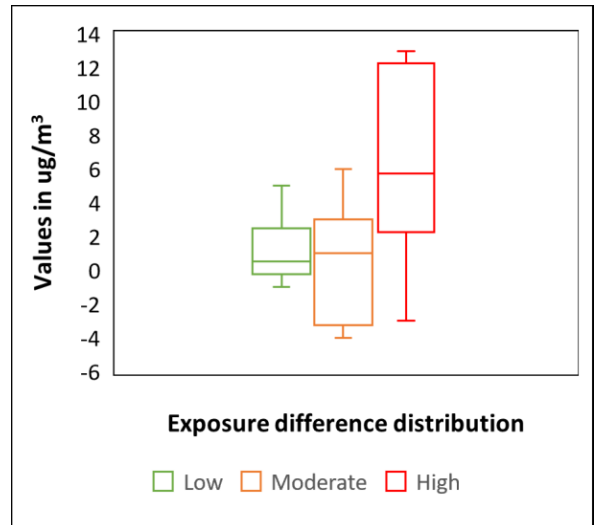

Figure 11: Box plots demonstrating exposure difference distribution with in the same category

In one of the WHO projects, i.e. HRAPIE (Health risks of air pollution in Europe), health impacts of air pollution were assessed. It was quantified that there is a relatively increased risk ($RR^1$ = 1.021) of bronchitis symptoms in children aged 5–14 years per 1 µg/m³ change in annual mean $NO_2$ concentration [51]. In another study, stronger associations between highway proximity and particles on the exhaled nitrogen oxide in children with asthma, showing increased inflammation of airway [49]. Based on these findings, we can say that adopting alternative routes that lie in the same exposure category i.e. but with some reduced exposure to $NO_2$ concentration can also bring health benefits for further 26 % of participants i.e. school children.

In most of the past studies, RR (relative risk) is estimated based on 10 μg/m³ increase of $NO_2$ [51,49,50,52,53]. On similar lines, an effort was made to quantify health impacts due to long term exposure of $NO_2$ in HRAPIE project. It was found that increases in $NO_2$ concentrations (per 10 μg/m³, annual means) are associated with increases in all-cause mortality with RR of 1.055 (95% CI: 1.031–1.080) [51]. A recent meta-analysis of 48 cohort studies conducted by Atkinson et al came up with the conclusion that there is a positive association between long term concentrations of $NO_2$ and all-cause mortality (RR=1.02 [95% CI: 1.01-1.03] per 10 μg/m$^3$ increment in $NO_2$). Similar association was found for cause-specific such as cardiovascular (RR = 1.03 [95% CI: 1.02, 1.05]), respiratory (RR=1.03 [95% CI: 1.01, 1.05]), and lung cancer mortality (RR=1.05 [95% CI: 1.02-1.08]) [54]. As mentioned earlier that in our study every 1 out of 3 participants (kids with escorting parent) have alternative routes available with reduced $NO_2$ annual concentrations of around 10 μg/m$^3$. So we can say that these kids and escorting parent can reduce the risks associated with exposure to high $NO_2$ concentrations by adopting the alternative routes. In total around 60 % (34% + 26%) of the participants (N= 62) have alternative routes available with reduced levels of $NO_2$ concentrations as compared to current routes.

---

[1] Relative risk is a ratio between the chance of occurrence of an event in the exposed group as compared to the non-exposed group. RR gives information about the higher and lower likelihood of an event between the exposure and the non-exposure group.

*4.3 Intervention Effectiveness*

In this section, some insights into the effectiveness of the implemented intervention are provided. The effectiveness is measured based on the responses of the questionnaire filled by the participants after the intervention. The feedback questionnaire asked the participants the following questions.

1. Did you learn something new by participating in this study? If you select YES, then please explain why?
2. Having learned about your exposure to air pollution, do you think you will change the way you escort your kid to school? Yes/No please explain
3. Do you think there is something you can do to reduce air pollution in your city? If you select YES, then please explain why?
4. Did you like the study? Rate from 1(hate) – 5 (love)

Around 88 % of the participants mentioned *YES* as the response to the 1$^{st}$ Question. The responses are encouraging as most of them reported that participation in the study increases their knowledge related to pollutant exposure and further about the commuting to school in a healthier way. From the 60% participants (N=62) that could bring some benefits by adopting alternative routes, 77 % of the participants (i.e. n=48) responded positively to the 2$^{nd}$ Question by reporting that they started following the suggested alternative routes to school with the least exposure to air pollutants. We further analyzed the reasoning for those 23 % participants who responded negatively to this question. 13% of them are those who also did not show agreement with the question asked about "environmental pollution in a locality is a concern" in their response to the introductory questionnaire. This means that there is a need for awareness campaigns to educate people about health impacts due to pollutant exposure. Another 10% participants were those who showed reluctance to follow the suggested alternative routes due to safety concerns. It is observed that the children of these participants are mostly from the age category of 9 -12 years old. This concern is probably because children started going to school independently at this age and parents are extra conscious about their safety. Some of the responses are provided as it is.

> "the alternative route is certainly greener but perhaps not safer for young cyclists due to lack of separate cycle path and footpath".
>
> "Because the alternative route is even more dangerous than the current one - first safety, then air pollution ....".

For question 3, 40 % of the participants showed good civic sense by responding positively that they can take certain actions to reduce air pollution in the city. A couple of respondents showed a commitment to increase cycle use and reduce car use as much as possible. Some participants have an intention to plant trees in their locality. One of the participants has a plan to encourage municipality about the improvement in walking/cycling paths. The escorting parents from age group below 40 years seem more enthusiastic and pro-environmental as most of them fall in the age category of 30 – 40 years, as mentioned in Table 1. Participants in overall like the study by giving a positive rating, the average score (from 104 participants) estimated is around 4.1 out of 5.

**5. Discussion**

In the last two decades, many environmental and transportation experts have explored the effect of traffic-related air pollution on active commuting. However, very few have tried to find the measures that commuters can proactively use to protect themselves from high pollutant exposure and its impact. This study examines alternative routes for active school commuting as one such measure and evaluates its potential. Based on the rules for selection of alternatives routes, as mentioned in section 2.2, 60 % of the participants have the availability of alternatives routes in which significant reduction of $NO_2$ exposure has been observed. This finding is consistent with those reported in other limited studies conducted to optimize routes for exposure reduction. A study for Greater Manchester, UK, was conducted in which the fastest home to school routes and routes with least $NO_2$ exposure were simulated. For nearly 60% of the walking routes, a less polluted route was found, based on cumulative $NO_2$ exposure [22]. A study was conducted in Montreal, Canada with the aim to calculate the shortest route (based on length) and the "cleanest route" based on $NO_2$ exposure. The Montreal study has found alternative routes for 57% of the cycling trips with a mean decrease in $NO_2$ concentration of 1.45 μg/m$^3$ [55]. The results of these studies are comparable to our study and it can be concluded that there is a significant potential to reduce school commuting exposure by suggesting alternative routes. In our study, we not only estimate this potential but also designed and implemented an informational intervention for its exploitation.

Several school travel behavior change initiatives have been taken to improve physical activity among children. These initiatives also reduce traffic congestion and pollution by discouraging car use. "Safe Route to School" (SRTS), "Walking School Bus", "Ride2School Program", "School

Travel Planning" (STP), "Bikeability" and "Beat the street" are among one of the prominent implemented programs targeted to promote physical activity in America (US), Australia, New Zealand, Canada and United Kingdom (UK) [56]. US government has spent around 1 billion dollars in the SRTS program to promote AST and the result was 4.5 % absolute increase in walking and cycling over the course of 5 years [57]. Ride2School initiative was taken in Australia with insignificant improvement in active school commuting of 2 % [15]. Similarly, STP was conducted in New Zealand and Canada with a modest increase of 2.1% and 1.7 % respectively in active school travel observed from the student reported follow-up data [13,16]. In UK a technology-based intervention "Beat the street" has implemented in school that resulted in 10 % increase in the AST in the intervention group[14]. The interventions mentioned above tried to influence children's school travel behavior by focusing on strategies like education, motivation, encouragement, and infrastructure improvement. All of these intervention provided the generic information without focusing the individual contextual information. It is found that intervention targeting an easy travel behavioral change are more effective provided that they are equipped with individual-specific feedback and benefits of change. Indimark® is found to be one of the successfully initiative implemented in Europe and Australia by providing tailored feedback to promote individual pro-environmental travel behavior. The reported estimate of travel behavior change is a maximum of 15 % which shows its impact [28]. Keeping in mind the similar past studies, we designed the intervention by combining multiple strategies such as education, encouragement, feedback (customized) and justification (benefits/relevance) to convince participants to adopt suggested school routes for pollutant exposure reduction. It is observed that the implemented intervention has a significant effect on participant school travel behavior. The reason behind such a positive impact is due to the high level of associated benefit. Furthermore, other reasons for its significant effect are the relevance of the intervention to the study area (pollution is a concern for the residents of the region) and targeting vulnerable group (i.e. parents/guardians are sensitive to their children health).

In addition to the above, the results of the interventions effectiveness results provide evidence that such study can be successfully managed remotely without any face to face connect by using apps (e.g. R2S online platform) and web tools. However, it is should be decided carefully based on the target audience. The concept and framework of this interventional study can be applied together with other studies in several ways. For example, in the short term, it can be used to assist the Route2school program in Belgium to select walking and cycling routes that address both safety and air quality concerns for children while school commuting. In the long term, this initiative can be coupled with other interventions set by cities to promote green and active commuting to improve accessibility, safety, and air quality for travelers in a holistic manner.

## 6. Conclusions

This paper describes the development, application and analysis of the air quality based informational intervention to promote healthy (less polluted routes) school commuting. The customized information is provided to each individual about the active school travel choices with the possible personal benefits achieved such as reduced pollutant exposure. It was found that around 34 % of the participants in the study group have the availability of walking/cycling routes with significantly less pollutant exposure (with a difference ≥ 10 µg/m$^3$ of $NO_2$ concentration) as compared to current routes. Further based on the responses in the feedback questionnaire, it was observed that for those participants where lower polluted alternatives are available, 77 % of such participants started following the suggested alternatives school routes. These findings provide evidence that this intervention can significantly promote school commuting focusing on least pollutant exposed routes. There is a clear potential for such soft interventions to be instigated on a wider scale with a correspondingly larger impact on individual health and wellbeing. As the benefits of reduction in pollutant exposure are considered in conjunction with the well-established health benefits of physical activity, such intervention based initiatives should be attractive to local and national governments. Based on the results of the large scale implementation, this intervention study can be further extended by developing a publicly available smartphone application that helps individuals to identify low pollutant walking/cycling routes which can help deliver health benefits for children and other individuals.

**Acknowledgements**

This project has received funding from the European Union Horizon 2020 research and innovation programme under grant agreement No 689954. This paper reflects the authors views. The European Commission is not liable for any use that may be made of the information contained therein. We would also like to thank Flemish Institute for Technological Research (VITO) for providing us with the running annual means of $NO_2$ concentration for entire Flanders at fine spatial resolution of 100 $m^2$.